\documentclass[pre,floats,superscriptaddress,usenames]{revtex4}
\usepackage[T1]{fontenc}
\usepackage[latin9]{inputenc}
\usepackage{amsmath}
\usepackage{amssymb}
\usepackage{graphicx}
\usepackage{esint}

\makeatletter

\usepackage{amssymb}
\usepackage{epsfig}

\usepackage{multirow}

\usepackage{amsmath}
\usepackage{graphicx}

\usepackage[dvipsnames]{color}

\newcommand{\be}{\begin{equation}}
\newcommand{\ee}{\end{equation}}
\newcommand{\bea}{\begin{eqnarray}}
\newcommand{\eea}{\end{eqnarray}}

\renewcommand{\ss}{{\pmb \sigma}}

\newcommand{\rr}{{\bf r}}

\newcommand{\vv}{{\bf v}}

\newcommand{\uu}{{\bf u}}
\newcommand{\FF}{{\bf F}}
\newcommand{\bk}{{\bf k}}
\begin{document}
\date{\today}


\title{Weighted density Lattice Boltzmann approach to fluids under confinement}

\author{Umberto Marini Bettolo Marconi $^{a}$ $^{\ast}$\thanks{$^\ast$Corresponding author. Email: umberto.marinibettolo@unicam.it
\vspace{6pt}}
and  Simone Melchionna  $^{b}$
\\\vspace{6pt}  $^{a}${ \em {Scuola di Scienze e Tecnologie, Universit\`a di Camerino, Via Madonna delle Carceri, 62032 ,
Camerino, Italy, INFN, sez. di Perugia, Italy }};
$^{b}$ { \em {CNR-IPCF, Consiglio Nazionale delle Ricerche, , Dipartimento di Fisica,  Universit\`a 
Sapienza, Piazzale A. Moro 2, 00185,
Roma, Italy }}\\\vspace{6pt}\received{v4.5 released September 2009} }

\begin{abstract}
The Enskog-like kinetic approach, recently introduced by us to study
strongly inhomogeneous fluids, is reconsidered in order to improve
the description of the transport coefficients. The approach is based
on a separation of the interaction between hydrodynamic and non-hydrodynamic
parts. The latter is treated within a simple relaxation approximation.
We show that, by considering the non-hydrodynamic part via a weighted
density approximation, we obtain a better prediction of the transport
coefficients. By virtue of the simplicity of the kinetic equation
we are able to solve numerically the phase space distribution in the
presence of inhomogeneities, such as confining surfaces, via a Lattice
Boltzmann method. Analytical estimates of the importance of these
corrections to the transport coefficients in bulk conditions is provided.
Poiseuille flow of the hard-sphere fluid confined between two parallel
smooth walls is studied and their pore-averaged properties are determined.
\end{abstract}


\maketitle

\section{Introduction}

The technological and industrial relevance of processes where molecular
fluids flow through channels of nanometric
diameter require to develop methods to control and
predict such phenomena \cite{Squires,Schoch,Bruus,Kirby,Bocquet,Bhatia}.
Among the various techniques available today, one can mention continuum-based
computational fluid dynamics methods, finite-element methods for Navier-Stokes
equations \cite{Karniadakis}, Molecular Dynamics \cite{BocquetBarrat,Hartkamp},
Kinetic theory \cite{Pozhar} Dynamic Boltzmann free-energy functional
theory \cite{Espanol}, and Dynamical Density Functional theory for
dense atomic liquids \cite{Archer,Melchionna2007}. In these methods,
one studies either the evolution of the local particle density $n(\rr,t)$
or the evolution of the one-particle phase-space density distribution
function. 

The strategy based on the solution of the phase-space distribution
function is known as the Lattice Boltzmann method \cite{LBgeneral,vergassolabenzisucci}
and is based on a discretized version of the Boltzmann transport equation,
originally conceived to solve at a reduced computational cost the
fluid dynamics equations. It can deal with complicated geometries
and is numerically robust but, conversely, is typically based on top-down
strategy: the phenomenological parameters, such
as the viscosity and the thermal conductivity, are utilized as lumped parameter
whose values are provided by some experiment or
macroscopic model. Since the underlying microscopic level is unexplored,
in principle one cannot infer the nature of the atomic constituents.
In addition, since the phenomenological transport coefficients are
fixed from bulk measurements, it is awkward to control how they change
under inhomogeneous conditions, such as those due to confinement in
narrow spaces.

In a series of papers we have considered an alternative approach 
\cite{Melchionna2008,Melchionna2009,UMBM2011},
that does not require the \emph{a priori} knowledge of the transport
coefficients, nor the equation of state, but directly tackles the Boltzmann
equation corresponding to a well defined microscopic model. A convenient
route is to consider a system of hard-spheres and possibily 
adding attractive potential tails. The Hamiltonian determines the
cross-section entering the Boltzmann-Enskog (BE) equation, thus establishing
a link between microscopic and macroscopic properties. Using such
bottom-up approach, it is possible to determine all transport coefficients
and the equation of state from the BE equation, without introducing
empirical ingredients. Such a strategy was pioneered by Luo, who
carried out a systematic derivation of the Lattice Boltzmann equation
starting from the Enskog equation. He obtained the equation of state
for non-ideal gases together with thermodynamic consistency \cite{Luo}.

The present method not only takes into account the non-local nature
of the collision operator and leads to a non-ideal gas equation of
state, but also reproduces the dependence of the transport coefficients
on the density. The approximation scheme utilizes concepts inspired
by the Density Functional Theory in the weighted density approximation
(WDA) version \cite{Tarazonawda,Evanswda}, as regards to the use
of a smoothed density field and the equilibrium pair distribution
function as key quantities. At equilibrium our approach gives the
same equations of the WDA and for such a reason we name it Weighted
Density Lattice Boltzmann method (WDLB). Finally, the method is employed
to derive a practical numerical scheme to study liquids flowing in confined
geometries.

The purpose of this article is also to improve the WDLB by considering
corrections to the viscosity and to the thermal conductivity. Such
corrections are already present in the Enskog theory and, since they
arise from the coupling between kinetic and collisional modes, they
are named cross-over contributions to the transport coefficients and
depend linearly with the fluid density. To this aim, we follow the
clever approximation introduced by Dufty, Santos and Brey (DSB) \cite{DSB}
to simplify the revised Enskog theory (RET) equation \cite{Beijeren},
without spoiling its good features. In particular, this approach preserves
the non-locality of interactions, it accounts for the static pair correlations
in a realistic way and provides a reasonable equation of state. In
essence, Brey and coworkers proposed to replace the complex collision
kernel featuring in the original RET (see eq. (\ref{RETcollision})
below) by a more tractable object. 

The idea behind this proposal is to separate the relevant slow hydrodynamic
modes from the fast evolving and less relevant kinetic modes. The
first ones are then carefully handled, whereas the latter are treated
by resorting to a simple relaxation time approximation, thus
avoiding the effort of computing difficult integrals involving high
velocity momenta of the collision operator. The resulting approximated
collision operator consists of two pieces, one acting on the hydrodynamic
modes and the second determining the relaxation towards equilibrium
of the kinetic modes through the simplest approximation of the Boltzmann
equation, the Bhatnagar-Gross-Krook (BGK) kernel \cite{BGK}. 

Through this simplification, one obtains a representation where
a set of effective fields describes the effect of intermolecular
interactions in a self-consistent fashion, and the resulting  transport equation
is amenable to
numerical solution. In a series of papers 
\cite{Lausanne2009,JCP2011,JCP2011B} we have numerically implemented
this approach \cite{DSB} by developing a numerical algorithm in the framework of
the Lattice Boltzmann method. Here, we consider
the effect of a more refined treatment of the relaxation term, as
proposed by Santos, Montanero, Dufty and Brey (SMDB) \cite{SMDB}
in order to account for the cross-over contributions to the transport
coefficients.

This paper is organized as follows: in section \ref{Model} we briefly
review the employed kinetic method in order to present the extension
of the WDLB approach, and motivate the introduction of the new terms
into the equation of evolution used in numerical applications. In section \ref{Specific} we give the explicit representation
of the collision operator in the case of hard spheres and compute
the effective fields. In section \ref{Analytic} we present results for the transport coefficients.
  Finally, in section \ref{Applications} we present
the numerical test of the new approximation. In section \ref{Conclusions}
we make some concluding remarks.

\section{Model}

\label{Model}

The evolution of the phase-space one-particle distribution $f(\rr,\vv,t)$
is represented by the following transport equation \cite{Kreuzer,DeGroot,McLennan}:
\begin{eqnarray}
\partial_{t}f(\rr,\vv,t)+v_{i}\partial_{i}f(\rr,\vv,t)+\frac{F_{i}(\rr)}{m}\frac{\partial}{\partial v_{i}}f(\rr,\vv,t)=\Omega[f](\rr,\vv,t)\label{uno}
\end{eqnarray}
where $\FF$ is an external force and $\Omega[f]$ represents the
effect of the interactions among the fluid particles which we shall
specify later. We adopt the Einstein summation convention that repeated
indices are implicitly summed over and the notation $\partial_{i}$
to indicate the partial derivative w.r.t. the i-th component of the
vector $\rr$ and $\partial_{t}$ to indicate the partial derivative
w.r.t. time.

In principle, after giving an explicit representation of $\Omega$
according to the underlying microscopic model, equation (\ref{uno}) can
be solved numerically. The case of non-local functionals $\Omega$,
such as the RET, is computationally demanding. Alternatively, one
can consider very simple forms of $\Omega$, the BGK being the
simplest \emph{ad hoc } choice. Our strategy considers a non-local
functional $\Omega$, but factorizes the velocity dependence in terms
of a local Maxwellian distribution times Hermite
tensor polynomials \cite{Lausanne2009}.
In addition, since one is chiefly interested in the evolution of the hydrodynamic
variables it is possible to further simplify the form of $\Omega$
without altering the local transfer of momentum and energy.

In this framework we implicitly assume that after few molecular collisions the system reaches
a state of local thermodynamic equilibrium characterized by the hydrodynamic
variables $n(\rr,t),\uu(\rr,t),T(\rr,t)$, being the number density,
local fluid velocity and local temperature respectively. We further
assume that the phase space distribution depends on space and
time only through these fields and their gradients.

The five hydrodynamic fields can be obtained from $f(\rr,\vv,t)$
through the relations 
\begin{equation}
\left(\begin{array}{ccc}
n(\rr,t)\\
n(\rr,t)\uu(\rr,t)\\
\frac{3}{2}k_{B}n(\rr,t)T(\rr,t)
\end{array}\right)=\int d\vv\left(\begin{array}{ccc}
1\\
\vv\\
\frac{m(\vv-\uu(\rr,t))^{2}}{2}
\end{array}\right)f(\rr,\vv,t).\label{colonna}
\end{equation}
 where $k_{B}$ is the Boltzmann constant.

We also introduce the corresponding moments of the collision operator

\begin{equation}
\left(\begin{array}{ccc}
0\\
{\bf C}(\rr,t)\\
Q(\rr,t)
\end{array}\right)=\int d\vv\left(\begin{array}{ccc}
1\\
m\vv\\
\frac{m(\vv-\uu(\rr,t))^{2}}{2}
\end{array}\right)\Omega(\rr,\vv,t).\label{colonna}
\end{equation}

By multiplying both sides of eq. (\ref{uno}) by the lowest Hermite
tensorial polynomials (that is, $1,$ $\vv$, and $m(\vv-\uu)^{2}/2$)
and integrating w.r.t. $\vv$ we obtain the following set of balance
equations for the five hydrodynamic variables:

\begin{equation}
\partial_{t}n(\rr,t)+\partial_{i}(n(\rr,t)u_{i}(\rr,t))=0\,,\label{continuity}
\end{equation}

\begin{equation}
mn(\rr,t)[\partial_{t}u_{j}(\rr,t)+u_{i}(\rr,t)\partial_{i}u_{j}(\rr,t)]+\partial_{i}\tilde{P}_{ij}(\rr,t)-F_{j}(\rr)n(\rr,t)-C_{j}(\rr,t)=0,\label{momentumcontinuity}
\end{equation}
 and 
\begin{equation}
\frac{3}{2}n(\rr,t)k_{B}[\partial_{t}T(\rr,t)+u_{i}(\rr,t)\partial_{i}T(\rr,t)]+\tilde{P}_{ij}(\rr,t)\partial_{i}u_{j}(\rr,t)+\partial_{i}\tilde{q}_{i}(\rr,t)-Q(\rr,t)=0\label{energyequation}
\end{equation}

The momentum and temperature equations (\ref{momentumcontinuity})
and (\ref{energyequation}) contain two additional quantities, a pressure
term 
\begin{equation}
\tilde{P}_{ij}(\rr,t)=m\int d\vv f(\rr,\vv,t)(\vv-\uu)_{i}(\vv-\uu)_{j}\label{pkin}
\end{equation}
 and a heat flux term: 
\begin{equation}
\tilde{q}_{i}(\rr,t)=\frac{m}{2}\int d\vv f(\rr,\vv,t)(\vv-\uu)^{2}(\vv-\uu)_{i}\label{qkin}
\end{equation}
 which in the $n\to0$ limit can be identified with the kinetic components
of the pressure tensor and heat flux, respectively. As we shall see in section \ref{Specific},
these two quantities also contain contributions which are neither
purely kinetic nor purely collisional.

One can also identify the last term in equation (\ref{momentumcontinuity})
with the gradient of the collisional contribution to the pressure tensor,
$P_{ji}^{c}$, 
\begin{eqnarray}
C_{i}(\rr,t) & = & -\frac{1}{m}\partial_{j}P_{ji}^{c}(\rr,t)\label{gradient}
\end{eqnarray}
and $Q$ with a combination of the gradient of the collisional contribution
to the heat flux vector $q_{i}^{c}$ and the pressure tensor times
the strain rate: 
\begin{eqnarray}
Q(\rr,t) & = & -\frac{1}{m}[\partial_{j}q_{j}^{c}(\rr,t)+P_{ji}^{c}(\rr,t)\partial_{j}u_{i}(\rr,t)].\nonumber \\
\label{gradienth}
\end{eqnarray}
The second equality in eqs. (\ref{gradient}) and (\ref{gradienth})
identifies the effective fields $C_{i}$ and $B$ with the
thermodynamic fields. 

Formally, we have derived a set of equations
for the five hydrodynamic variables, but these do not form a closed
system, because we still need either to specify the non-equilibrium
distribution function $f$ or to assume some phenomenological constitutive
relations between the gradients of $P_{ij}$ and $q$ (the sums of
the kinetic and collisional parts) and the fluxes. Using the
latter strategy one obtains the Navier-Stokes equations \cite{Batchelor}.

The alternative route is to focus on the evolution of $f$, with the collision
operator $\Omega$ being  in general a non-linear functional of $f$,
so that any kind of analytic work results hard. The simplest
and crudest way to proceed is to approximate $\Omega$ by the following
relaxation ansatz as done by BGK \cite{BGK}
\begin{equation}
\Omega^{BGK}(\rr,\vv,t)=-\omega_{0}[f(\rr,\vv,t)-n(\rr,t)\phi_{M}(\rr,\vv,t)],\label{relaxationtime}
\end{equation}
where 
\[
\phi_{M}(\rr,\vv,t)=[\frac{m}{2\pi k_{B}T(\rr,t)}]^{3/2}\exp\left(-\frac{m(\vv-\uu)^{2}}{2k_{B}T(\rr,t)}\right)
\]
is the local Maxwellian. The BGK recipe for $\Omega$ preserves the
number of particles, the momentum and the kinetic energy, in other
words it fulfills the physical symmetries and conservation laws of
the fluid. Finally  it gives a quite uninteresting thermodynamic behavior
corresponding to an ideal gas. Instead, it gives non vanishing values
of the transport coefficients, which turn out to be functions of phenomenological
parameter $\omega_{0}$ \cite{BGK}. 

In view of the interest for the non-ideal gas behavior one seek for a
better approximation capable of predicting a non-trivial equation
of state and transport coefficients. The Boltzmann equation for hard
spheres has non-trivial transport coefficients but lacks a satisfactory
equation of state. The RET form of $\Omega$ has the sought features,
but is hardly tractable in numerical work \cite{Beijeren}. 

An interesting and practical
reduction of the RET collision operator has been devised by Dufty
et \emph{al}. \cite{DSB}. Their idea is to  separate the contributions of $\Omega$
to the hydrodynamic equations from those affecting the evolution of the
non-hydrodynamic modes, by projecting the collision term onto the
hydrodynamic subspace spanned by the functions $\{1,\vv,v^{2}\}$
and onto the complementary kinetic subspace: 
\begin{equation}
\Omega={\cal P}_{hydro}\Omega+(I-{\cal P}_{hydro})\Omega\label{splitting}
\end{equation}
 with 
\begin{equation}
{\cal P}_{hydro}\Omega=\frac{1}{k_{B}T(\rr,t)}\phi_{M}(\rr,t)\Bigl[(\vv-\uu)\cdot{\bf C}(\rr,t)+[\frac{m(\vv-\uu(\rr,t))^{2}}{3k_{B}T}-1]Q(\rr,t)\Bigr] .
\end{equation}
The second term, $\Omega_{\perp}\equiv(I-{\cal P}_{hydro})\Omega$ is further approximated
as: 
\begin{equation}
\Omega_{\perp}\approx-\omega_{0}[f(\rr,\vv,t)-n(\rr,t)\phi_{M}(\rr,\vv,t)]  \label{dufty1}
\end{equation}
and (\ref{uno}) by a simpler equation, where the complicated interaction
between non hydrodynamic modes is approximated via a BGK-like relaxation
term  with $\omega_{0}$
being a phenomenological collision frequency chosen as to reproduce
the kinetic contribution to the viscosity.

It was pointed out by Santos et al. \cite{SMDB} that better results
for the transport properties could be obtained upon modifying the
BGK relaxation term in (\ref{dufty1})  as to allow
the relaxation time to depend on the spatial inhomogeneities of the
local equilibrium state. This effect depends on the collisional transfer
mechanism associated with the difference in position of the colliding
particles. To achieve this goal, following Santos, Montanero, Dufty
and Brey  \cite{SMDB}, we complement the relaxation term in
eq. (\ref{dufty1}) in the following way: 
\begin{eqnarray}
\Omega_{\perp}^{SMDB}(\rr,t) & = & -\omega_{0}[f(\rr,\vv,t)-n(\rr,t)\phi_{M}(\rr,\vv,t)]+(m\beta)^{2}\phi_{M}(\rr,\vv,t)\times\nonumber \\
 &  & \Bigl\{\left[(\vv-\uu)_{i}(\vv-\uu)_{j}-\frac{1}{3}(\vv-\uu)^{2}\delta_{ij}\right]\frac{1}{2}A_{ij}(\rr,t)+\nonumber \\
 &  & +\left[\frac{m\beta(\vv-\uu)^{2}}{2}-\frac{5}{2}\right](\vv-\uu)_{i}\frac{1}{5}B_{i}(\rr,t)\Bigr\}\label{eq:dufty2}
\end{eqnarray}
 The extra terms in eq. (\ref{eq:dufty2}) give a better representation
of the decays towards equilibrium as discussed in ref. \cite{SMDB}. The first
term is proportional to $\omega_{0}$, which assumes
a constant value throughout the system, whereas the second term may
depend on the spatial inhomogeneities of the velocity
and of the temperature fields. Such a dependence occurs through the
functions $A_{ij}(\rr,t)$ and $B_{i}(\rr,t)$ and thus is modulated
by the spatial structure of the fluid. In the case of a uniform system
the relaxation process may occur only via the term proportional to
$\omega_{0}$, when the distribution function $f$ is different from
the local equilibrium distribution. The inclusion of these terms also
yields a quantitative improvement of the transport coefficients with
respect the simpler theory discussed previously. 

The new terms can be obtained by computing the following projections
of the collision operator: 
\begin{equation}
A_{ij}(\rr,t)=\int d\vv[(\vv-\uu)_{i}(\vv-\uu)_{j}-\frac{1}{3}(\vv-\uu)^{2}\delta_{ij}]\,\Omega(\vv,\rr,t)\label{dij}
\end{equation}
 and 
\begin{equation}
B_{i}(\rr,t)=\int d\vv[(\vv-\uu)^{2}-5\frac{k_{B}T}{m}](\vv-\uu)_{i}\,\Omega(\vv,\rr,t)\label{si}
\end{equation}

By construction, the new terms do not affect the structure of the
continuity equation (\ref{continuity}), momentum balance (\ref{momentumcontinuity})
and energy balance equations (\ref{energyequation}). The SMDB equation
can be cast in the form:
\begin{eqnarray}
\partial_{t}f(\rr,\vv,t) & + & v_{i}\partial_{i}f(\rr,\vv,t)+\frac{F_{i}(\rr)}{m}\frac{\partial}{\partial v_{i}}f(\rr,\vv,t)=\Omega_{\perp}^{SMDB}(\rr,t)\nonumber \\
 & + & m\beta\phi_{M}(\rr,\vv,t)\Bigl\{(\vv-\uu)\cdot{\bf C}(\rr,t)+(\frac{m\beta(\vv-\uu)^{2}}{3}-1)Q(\rr,t)\Bigr\}\label{SMDB}
\end{eqnarray}
where $\beta=1/k_{B}T(\rr,t)$. A simple calculation allows to compute
the deviation $\delta f$ of the distribution function $f(\rr,\vv,t)$
form its local equilibrium value $f_{loc}=n(\rr,t)\phi_{M}(\rr,t)$.
We replace $f\to f_{loc}$ in the left hand side of eq. (\ref{SMDB})
and obtain: 
\begin{eqnarray}
 &  & \Bigl\{\bigl[\partial_{t}n+\partial_{i}(nu_{i})\bigr]+\bigl[n\partial_{t}u_{k}+nu_{i}\partial_{i}(nu_{k})+\partial_{k}(nT)-\frac{F_{k}}{m}-C_{k}\bigr]\frac{m(v_{k}-u_{k})}{k_{B}T}\nonumber \\
 &  & +\frac{n}{2T}\bigl[\partial_{t}T+u_{i}\partial_{i}T+\frac{2}{3}T\partial_{i}u_{i}-\frac{2}{3n}Q\bigr](\frac{m(\vv-\uu)^{2}}{k_{B}T}-3)+[\frac{n}{T}\partial_{i}T-\frac{m\beta}{5}B_{i}]\bigl(m\frac{(\vv-\uu)^{2}}{2k_{B}T}-\frac{5}{2}\bigr)(v_{i}-u_{i})\nonumber \\
 &  & +m\beta\,[n\,\partial_{i}u_{k}-\frac{m\beta}{2}A_{ik}]\Bigl((v_{i}-u_{i})(v_{k}-u_{k})-\frac{(\vv-\uu)^{2}}{3}\delta_{ij}\Bigr)\Bigr\}\phi_{M}(\rr,\vv,t)=-\omega_{0}\delta f(\rr,\vv,t)
\end{eqnarray}
Since the projections of $\delta f$ over ($1,(v_{i}-u_{i}),m(\vv-\uu)^{2}/(k_{B}T)-3)$)
vanish , we impose that also the first three terms in the l.h.s. vanish.
These are the so-called solvability conditions and are precisely the
balance equations (\ref{continuity}), (\ref{momentumcontinuity})
and (\ref{energyequation}). We thus obtain the following explicit
representation of $\delta f$ 
\begin{eqnarray}
\delta f & = & -\frac{1}{\omega_{0}}\phi_{M}(\rr,\vv,t)\Bigl\{\Bigl[\frac{n(\rr,t)}{T}\partial_{i}T(\rr,t)-\frac{(m\beta)^{2}}{5}B_{i}\Bigr]\Bigl(\frac{m}{2}\frac{(\vv-\uu)^{2}}{k_{B}T(\rr,t)}-\frac{5}{2}\Bigr)(v_{i}-u_{i})\nonumber \\
 & + & m\beta\,\Bigr[n(\rr,t)\,\partial_{i}u_{k}(\rr,t)-\frac{m\beta}{2}A_{ik}\Bigr]\left((v_{i}-u_{i})(v_{k}-u_{k})-\frac{(\vv-\uu)^{2}}{3}\delta_{ik}\right)\Bigr\}\label{perturbed}
\end{eqnarray}
The fields $A_{ij}$ and $B_{i}$ represent a measure of the distortion
of $f(\rr,\vv,t)$ with respect to the local equilibrium $f_{loc}(\rr,\vv,t)$.
Using such an approximation for $\delta f$, we can now estimate how
the distortion affects the components of the pressure tensor (\ref{pkin}):
\begin{equation}
\tilde{P}_{ij}(\rr,t)=k_{B}T(\rr,t)n(\rr,t)\Bigl\{\delta_{ij}-\frac{1}{\omega_{0}}\Bigl((\partial_{i}u_{j}(\rr,t)+\partial_{j}u_{i}(\rr,t))-\frac{2}{3}\delta_{ij}\sum_{k}\partial_{k}u_{k}(\rr,t)\Bigr)\Bigl\}+\delta P_{ij}(\rr,t)\label{pexplicit}
\end{equation}
where the first term is the BGK result and 
\begin{equation}
\delta P_{ij}(\rr,t)=\frac{m}{\omega_{0}}\Bigl[A_{ij}(\rr,t)+A_{ji}(\rr,t)-\frac{2}{3}\delta_{ij}\sum_{k}A_{kk}(\rr,t)\Bigr]\label{extrapressure}
\end{equation}
stems from the $A_{ij}$ contribution. The heat flux is obtained by
substituting $\delta f$ in eq.(\ref{qkin}) 
\begin{equation}
\tilde{q}_{i}(\rr,t)=-\frac{5}{2}\frac{1}{m\omega_{0}}n(\rr,t)k_{B}^{2}T(\rr,t)\partial_{i}T(\rr,t)+\delta q_{i}(\rr,t)\label{qexplicit}
\end{equation}
with $\delta q_{i}(\rr,t)=\frac{1}{\omega_{0}}\frac{m}{2}B_{i}.$

It is worth noting that Luo \cite{LuoPRE} compared the approaches derived
from the Enskog equation with those based on the simplifying assumption
that the intermolecular forces are assimilated to a forcing self-consistent
term. He concluded that the latter are plagued by an inconsistency.
In fact, the pressure featuring in the momentum equation must also
feature in the energy equation, a condition which is not satisfied
if one treats the interactions in a mean-field fashion \cite{ShanChen,Swift}.
In this respect the present treatment has the correct form, since
eq. (\ref{energyequation}) contains the viscous dissipation term
representing the irreversible transformation of kinetic energy
into heat.

\section{Hard sphere model}

\label{Specific} 

At this stage we are interested in obtaining an explicit representation
of the fields $(C_{i},Q,A_{ij},B_{i})$ in order to determine the
viscosity and thermal conductivity, and to implement the numerical
solution of the evolution equation (\ref{SMDB}) by lattice methods.
To this purpose we use as a benchmark the prototypical model of dense
fluids, namely the hard-sphere system at moderate packing fractions. 
The RET collision operator involves only the single-particle phase
space distribution and accounts for static pair correlations, but
not velocity correlations. At high density it becomes more inaccurate
because  it disregards
the occurrence of correlated collisions. Nevertheless,
the RET approach is very accurate for our purposes and has the interesting
feature of taking into account the instantaneous transfer of energy
and momentum between particles whose distance equals the hard sphere
diameter. For this reason even to a spatially uniform state corresponds
a non-ideal gas pressure.

The RET collision operator reads \cite{Beijeren}: 
\begin{eqnarray}
\Omega^{RET}(\rr,\vv_{1},t) & = & \sigma^{d-1}\int d\vv_{2}\int d\hat{k}\Theta(\hat{{\bf k}}\cdot\vv_{12})(\hat{{\bf k}}\cdot\vv_{12})\times\nonumber \\
 &  & \left\{ g_{2}(\rr,\rr-\hat{{\bf k}}\sigma,t)f(\rr,\vv_{1}',t)f(\rr-\hat{{\bf k }}\sigma,\vv_{2}',t)-\right.\nonumber \\
 &  & \left.g_{2}(\rr,\rr+\hat{{\bf k}}\sigma,t)f(\rr,\vv_{1},t)f(\rr+\hat{{\bf k}}\sigma,\vv_{2},t)\right\} \label{RETcollision}
\end{eqnarray}
where $\vv_{1}'$ and $\vv_{2}'$ are scattered velocities determined
from $\vv_{1}'=\vv_{1}-(\hat{{\bf k}}\cdot\vv_{12})\hat{{{\bf k}}}$
and $\vv_{2}'=\vv_{2}+(\hat{{\bf k}}\cdot\vv_{12})\hat{{\bf k}}$,
$\sigma$ is the hard-sphere diameter, $\hat{{\bf k}}$ is the unit
vector directed from one particle to another, $\Theta$ is the Heaviside
function, and $g_{2}(\rr_{1},\rr_{2},t)$ is the pair correlation
function evaluated at contact distance $|\rr-\rr'|=\sigma$. It describes
the positional pair correlations corresponding to the inhomogeneous
single particle density profile $n(\rr,t)$. There is no velocity
dependence in the two particle correlation. 

Following Ref. \cite{MarconiTarazona}
we assume that $g_{2}$ at time $t$ is the same as the one corresponding
to the same system having a density profile equal to $n(\rr,t)$ at
thermodynamic equilibrium, an assumption often referred to as the adiabatic
approximation. Moreover, since the exact inhomogeneous form of $g_{2}$
is unknown, we resort to the Fischer-Methfessel prescription \cite{FischerMethfessel},
stating that $g_{2}(\rr_{1},\rr_{2},t)$ is given by the pair correlation
function at contact of an homogeneous system evaluated replacing $n$
by the following smeared density: 
\begin{equation}
\bar{n}(\rr,t)\equiv\frac{6}{\pi\sigma^{3}}\int_{|\rr'|<\sigma/2}d\rr'n(\rr+\rr')\,.
\end{equation}

Because we have access only to the first lower velocity moments of $f$,
we cannot solve the full problem (\ref{uno}) using such a
collision operator. The DSB and SMDB recipes represent viable options.
Recently we have used the SDB equation in conjunction with the Lattice
Boltzmann method to study one component, two component and ternary
mixtures with interesting applications also to electrolyte solutions
\cite{EPL2011,Lee2012}. Hereafter, we want to test the SMDB equation
implemented by an appropriate LB method and compare the results. 

The expressions of ${\bf C},B$ have been obtained before \cite{Melchionna2009}
and read 
\begin{eqnarray}
 &  & C_{i}(\rr,t)=-k_{B}T(\rr,t)n(\rr,t)\sigma^{2}\int d\hat{k}k_{i}g_{2}(\rr,\rr+\sigma\hat{{\bf k}},t|n)n(\rr+\sigma\hat{{\bf k}},t)\nonumber \\
 &  & \bigg[1-\frac{2}{\sqrt{\pi k_{B}T(\rr,t)/m}}\hat{k}\cdot[\uu(\rr+\sigma\hat{k},t)-\uu(\rr,t)]+\frac{T(\rr+\sigma\hat{k},t)-T(\rr,t)}{2T(\rr,t)}\bigg]\label{campoc}
\end{eqnarray}
and
\begin{eqnarray}
Q(\rr,t)={k_{B}T(\rr,t)}n(\rr,t)\sigma^{2}\int d\hat{k}g_{2}(\rr,\rr+\sigma\hat{{\bf k}},t|n)n(\rr+\sigma\hat{{\bf k}},t)\nonumber\\
\bigg[-\frac{\hat{k}\cdot[\uu(\rr+\ss\hat{k},t)-\uu(\rr,t)]}{2}+\frac{1}{\sqrt{\pi}}\sqrt{\frac{k_{B}T(\rr,t)}{m}}\frac{T(\rr+\sigma\hat{k},t)-T(\rr,t)}{T(\rr,t)}\bigg]
\label{campob1}
\end{eqnarray}

A similar calculation was performed to evaluate $A_{ij}$ and $B_{i}$
using (\ref{dij}) and (\ref{si}), respectively, under the approximation
that $\Omega^{RET}[f,f]$ featuring in these two equations was replaced
by $\Omega^{RET}[f_{loc},f_{loc}]$ \cite{Melchionna2007,Longuet}.
We find: 
\begin{eqnarray}
 &  & A_{ij}(\rr,t)=-\frac{k_{B}T(\rr,t)}{5m}n(\rr,t)\sigma^{2}\int d\hat{k}g_{2}(\rr,\rr+\sigma\hat{{\bf k}},t|n)n(\rr+\sigma\hat{{\bf k}},t)\nonumber \\
 &  & \bigg[\hat{k}_{j}[u_{i}(\rr+\sigma\hat{{\bf k}},t)-u_{i}(\rr,t)]+\hat{k}_{i}[u_{j}(\rr+\sigma\hat{{\bf k}},t)-u_{j}(\rr,t)]\bigg]
. \label{campod}
\end{eqnarray}
 Similarly, we  derive the correction to the heat flux vector 
\begin{equation}
B_{i}(\rr,t)=-2\frac{k_{B}^{2}}{m^{2}}T(\rr,t)n(\rr,t)\sigma^{2}\int d\hat{k}\hat{k}_{i}g_{2}(\rr,\rr+\sigma\hat{{\bf k}},t|n)n(\rr+\sigma\hat{{\bf k}},t)\bigg[T(\rr+\sigma\hat{{\bf k}},t)-T(\rr,t)\bigg].
\label{campos}
\end{equation}
 As shown in ref. \cite{JCP2011} a useful representation of the term
(\ref{campoc}) is obtained by decomposing it in the following way:
${\bf C}(\rr,t)=n(\rr,t)\left(\FF^{mf}(\rr,t)+\FF^{viscous}(\rr,t)\right)$,
where we can identify a force acting on the particle at position $\rr$
as due to the influence of all remaining particles in the system,
the so called potential of mean force \cite{HansenMcDonald}:
\begin{equation}
F_{i}^{mf}(\rr,t)=-k_{B}T\sigma^{2}\int d\hat{k}k_{i}g_{2}(\rr,\rr+\sigma\bk,t)n(\rr+\sigma\bk,t)
\end{equation}
 and a viscous term 
\begin{equation}
F_{i}^{viscous}(\rr,t)=2\sigma^{2}\sqrt{\frac{mk_{B}T}{\pi}}\int d\hat{k}k_{i}k_{j}g_{2}(\rr,\rr+\sigma\bk,t)n(\rr+\sigma\bk,t)(u_{j}(\rr+\sigma\bk)-u_{j}(\rr))
\end{equation}
 and a force due to the presence of thermal gradients 
\begin{equation}
F_{i}^{T}(\rr,t)=-\frac{\sigma^{2}}{2}\int d\hat{k}k_{i}g_{2}(\rr,\rr+\sigma\bk,t)n(\rr+\sigma\bk,t)k_{B}[T(\rr+\sigma\bk,t)-T(\rr,t)],\label{thermalforce}
\end{equation}

One can see that the form of the fields $C_{i},Q,A_{ij},B_{i}$ depends
on density, velocity profile and temperature. Only one contribution
to $C_{i}$ inherits the property of the density functional theory
of being fully determined by the density only. The other terms require
the knowledge of the hydrodynamic fields even in the present approximation,
since they are not slaved to the density field as it occurs in 
Dynamical Density Functional theory.

\section{Analytic estimate of the bulk transport coefficients from WDLB}
\label{Analytic}

Let us consider the limit of small velocity gradients and constant
density and temperature, so that:

\begin{equation}
A_{ij}(\rr,t)\simeq-\frac{k_{B}T}{m}n^{2}g_{2}(\sigma^{+})\frac{4\pi}{15}\sigma^{3}\bigg[\partial_{j}u_{i}+\partial_{i}u_{j}\bigg]\label{c1expl2}
\end{equation}
and, using eq.(\ref{extrapressure}):
\begin{equation}
\delta P_{ij}(\rr,t)=-\frac{1}{\omega_{0}}k_{B}Tn^{2}g_{2}(\sigma^{+})\frac{8\pi}{15}\sigma^{3}\bigg[\partial_{j}u_{i}+\partial_{i}u_{j}-\frac{2}{3}\delta_{ij}\sum_{k}\partial_{k}u_{k}\bigg]
\end{equation}

Let us consider a system of uniform density and temperature, and a
small sinusoidal velocity perturbation varying in a direction normal
to the streamlines:

\begin{equation}
\uu(x,t)=(0,u_{y}(0,t)\, e^{iq_{x}x},0)
\end{equation}
 By neglecting the non-linear term, we obtain from eq. (\ref{momentumcontinuity})
\begin{equation}
\frac{\partial}{\partial t}u_{y}(x,t)=-\frac{\partial}{\partial x}\tilde{P}_{xy}(x,t)+C_{y}(x,t)
\end{equation}
 and, substituting the expressions for $C_{y}$ and $\tilde{P}_{xy}$,
\begin{equation}
n\frac{\partial}{\partial t}u_{y}(x,t)=\Bigl[\frac{1}{\omega_{0}}k_{B}Tn\left(1+\frac{8\pi}{15}n\sigma^{3}g_{2}(\sigma^{+})\right)+\frac{4}{15}\sqrt{m\pi k_{B}T}n\sigma^{4}g_{2}(\sigma^{+})\Bigr]\frac{\partial^{2}u_{y}(x,t)}{\partial x^{2}}\label{treotto}
\end{equation}
The solutions have exponential form, $u_{y}(x,t)=u_{y}(0,t)e^{iq_{x}x}e^{-(\eta/n)q_{x}^{2}t}$,
where the shear viscosity $\eta$ is given by the expression inside square
brackets in the r.h.s. of eq. (\ref{treotto}).

The first term arises from the reduction of the velocity gradients
due to the displacement of particles from a region characterized by
a higher fluid velocity $\uu$ to a region of lower velocity and vice
versa. The third term takes into account the fact that neighboring
volume elements moving at different velocities can exchange momentum
due to the interaction of their surface molecules without involving
particle displacement. The second term $\eta^{(1)}$, called cross-over
term, has a mixed nature \cite{ChapmanCowling}. Finally, we set the collision
frequency to the value $\omega_{0}=\frac{16}{5}n\sigma^{2}g_{2}(\sigma)\sqrt{\pi k_{B}T/m}$
\cite{SMDB} giving the correct low density shear viscosity of
hard spheres of diameter $\sigma$ 
\begin{equation}
\eta=\frac{nk_{B}T}{\omega_{0}}\bigl[1+\frac{8\pi}{15}n\sigma^{3}g_{2}(\sigma^{+})\bigl]+\frac{4}{15}\sqrt{m\pi k_{B}T}n\sigma^{4}g_{2}(\sigma^{+}) .
\end{equation}

\subsection{Bulk viscosity}

In order to derive the expression for the bulk viscosity, $\eta_{b}$,
we consider a velocity field of the form $\uu=(u_{x}(x,t),0,0)$,
use the macroscopic relation 
\begin{equation}\frac{\partial P_{xx}(x,t)}{\partial x}=-(\eta_{b}+\frac{4}{3}\eta)\frac{\partial^{2}u_{x}(x,t)}{\partial x^{2}}
\end{equation}
to define $\eta_{b}$ and obtain after a simple calculation 
\begin{equation}
\frac{\partial P_{xx}(x,t)}{\partial x}=-\Bigl[\frac{4}{3}\frac{1}{\omega_{0}}k_{B}Tn\left(1+\frac{8\pi}{15}n\sigma^{3}g_{2}(\sigma^{+})\right)+\frac{4}{5}\sqrt{\pi}\sqrt{mk_{B}T}n\sigma^{4}g_{2}(\sigma^{+})\Bigr]\frac{\partial^{2}u_{x}(x,t)}{\partial x^{2}} .
\end{equation}
By comparing, the two expressions for the derivative of the pressure
tensor, we find $\eta_{b}=\frac{5}{3}\eta$ and conclude that $\eta_{b}$
is purely collisional with no kinetic and cross contributions.

\subsection{Thermal conductivity}

The heat flux correction $B_{i}$ for small gradients can be approximated
as 
\begin{equation}
\delta\tilde{q}_{x}(\rr,t)\simeq-\frac{1}{\omega_{0}}\frac{k_{B}^{2}}{m}T(\rr,t)n^{2}g_{2}(\sigma^{+})\frac{4\pi}{6}\sigma^{3}\frac{\partial T}{\partial x}\label{c1exp22}
\end{equation}
Let us consider the heat flux in a system at rest and of uniform density.
According to equation (\ref{energyequation}) 
\begin{equation}
\frac{3}{2}k_{B}n(\rr,t) \partial_{t}T(\rr,t)=
   \Bigl\{
   \frac{5}{2}\frac{1}{m\omega_{0}}n(\rr,t)k_{B}^{2}T
\Bigl[
       1+ng_{2}(\sigma^{+})\frac{4\pi}{15}\sigma^{3}
\Bigr]
    +\frac{2}{3}\sqrt{m\pi k_{B}T}g_{2}(\sigma^{+})n^{2}\sigma^{4}\frac{k_{B}}{m}
    \Bigl\}
\frac{\partial^{2}T}{\partial x^{2}}\label{thermalcond}
\end{equation}
yielding the thermal conductivity, $\lambda$, as the sum of the
three terms inside the parenthesis in the r.h.s. of (\ref{thermalcond}).
These are the kinetic, cross-over and collisional contributions to
$\lambda$, respectively. For the sake of comparison the Enskog formulae
\cite{ChapmanCowling} are 
\begin{equation}
\eta_{E}=\eta^{(0)}[1.016/g_{2}(\sigma^{+})+0.8bn+0.761g_{2}(\sigma^{+})b^{2}n^{2}]\label{etacc}
\end{equation}
 and 
\begin{equation}
\lambda_{E}=\lambda^{(0)}[1.025/g_{2}(\sigma^{+})+1.230\, bn+0.7764g_{2}(\sigma^{+})b^{2}n^{2}]\label{lambdacc}
\end{equation}
 where $b=2\pi/3\sigma^{3}$ and $\eta^{(0)}=\frac{5}{16\sigma^{2}}\sqrt{\frac{mk_{B}T}{\pi}}$
and $\lambda^{(0)}=\frac{75}{64\sigma^{2}}\sqrt{\frac{k_{B}^{3}T}{m\pi}}$.

We shall derive such corrections and compute numerically their effect
by using the SMDB approach.

\section{Numerical results}

\label{Applications} 

By using the results derived above, with the effective fields given by
(\ref{campoc}), (\ref{campob1}), (\ref{campod}) and (\ref{campos}),
we construct the corresponding Lattice Boltzmann algorithm to solve the kinetic equation
(\ref{SMDB}) on  a three-dimensional cubic lattice with lattice
constant $a$, assumed to be a fraction of the hard sphere diameter
$\sigma$. 

The continuous dependence of $f(\rr,\vv,t)$ on its argument $\rr$ 
is replaced by a finite set of nodes of coordinates $\rr_{i}$ of
a regular lattice spanning the physical space available to the particles.
The $\vv$ dependence is given by spanning $f$ over an orthonormal
basis formed by a finite number of tensorial Hermite polynomials.
We consider the so-called D3Q19 version, where the $\vv$-space is
characterized by $19$ discrete velocities. The state of the system
is specified by the discrete set of distribution functions $f_{\alpha}(\rr_{i},t)$ and
for the sake of simplicity, we only consider isothermal systems
and neglect the evolution of the $T$ field. The finer details of
the method have been described in refs. \cite{Melchionna2008,Melchionna2009}
and will not be repeated here. In the present implementation, the
method only contains the two additional terms proportional to $A_{ij}$
and $B_{i}$. 

The LB algorithm performs the following basic operations for every
mesh node: 
\begin{enumerate}
\item the discretized distribution is initialized by specifying the local
values of the density and fluid velocity;
\item the values of the fields $A_{ij},B_{i},C_{i},Q$ are computed via
eqs. (\ref{campoc})-(\ref{campos}) using the fields $\uu$ and $n$;
\item the distribution is evolved in time though the collision and streaming
steps via an explicit trapezoidal rule;
\item the new fields $n(\rr,t)$ and $\uu(\rr,t)$ are computed using the
evolved distribution function;
\item the cycle is repeated starting from step (2). 
\end{enumerate}
In this scheme, one relevant question pertains the choice of the BGK
relaxation frequency $\omega_{0}$. According to the Enskog picture,
the kinetic contribution to viscosity should be set equal 
to $\eta_{K}=\frac{5}{16\sigma^2}\sqrt{\frac{m k_B T}{\pi}}$  
and therefore $\omega_{0}=\frac{nk_{B}T}{\eta_{K}}$. In addition,
the low-density regime should be characterized by a constant dynamic
viscosity, as first observed by Maxwell and, as $n\rightarrow0$,
the viscosity should go to zero. This rather large baseline for viscosity
necessarily implies a strong departure from the local equilibrium.
Such a condition is numerically critical, as it corresponds to a mean free
path of the order of $\sigma$, whereas numerical stability imposes
that the mean free path should be of the order of the mesh spacing
$a$. In this case, in fact, the BGK term becomes effective in tethering
the distribution close to the local equilibrium form, coherently with
the small-Knudsen conditions employed in the theory developed so far.
For this reason, we choose a smaller value of the kinetic viscosity.
It is important to remember that also the cross-over term depends
on $\omega_{0}$, but nevertheless we can still determine its effect
on the global viscosity.

\begin{figure}[htb]
\begin{centering}
\includegraphics[clip,scale=0.6]{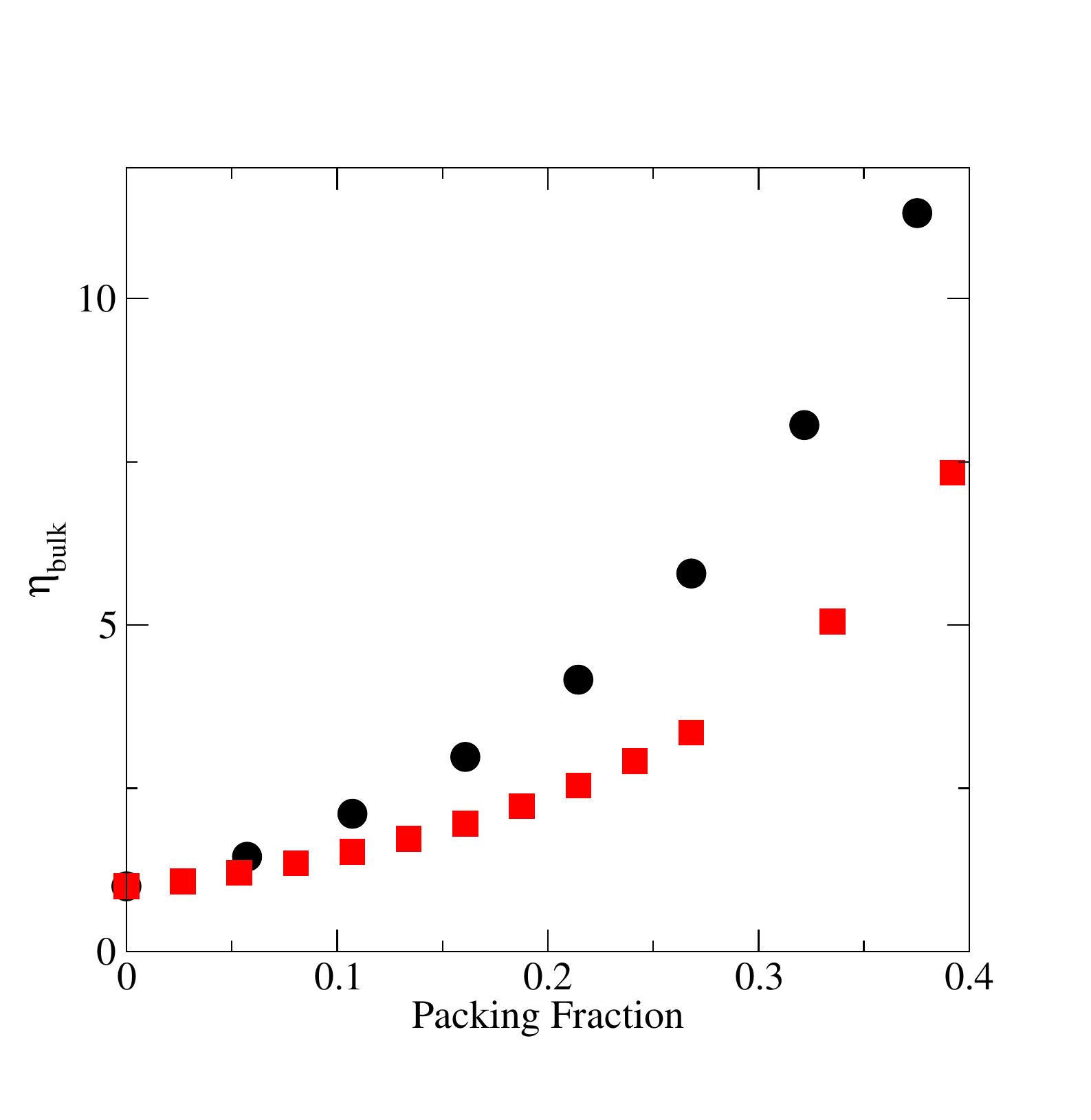}
 \caption{Shear viscosity computed in bulk conditions $\eta_{bulk}$ as a function
of the packing fraction (this quantity should not be confused with the bulk viscosity of Section 4.1).
Circles and squares are with and without
the cross-over correction to viscosity, respectively. }
\label{viscositybulk} 
\end{centering}
\end{figure}

In the numerical benchmark, we evaluate the effect of the cross-over
contributions to shear viscosity stemming from eq. (\ref{eq:dufty2}).
In this test we choose a hard-sphere diameter $\sigma=8a$ and the
packing fraction $\chi=\frac{\pi\sigma^{3}n}{6}$ ranges from dilute
gas to dense liquid conditions ($\chi\simeq0.40$). In Fig. \ref{viscositybulk}
we report the shear viscosity in bulk conditions as a function of the packing fraction $\chi$. 
 The measurements were performed by initially  perturbing a uniform quiescient state
by a sine wave of the type $u_z^0 \sin(q_x x)$ and monitored
that its relaxation occurred with a characteristic time $\tau(q_x)= \frac{\eta}{n_0} q_x^2$,
where $\eta$ is the shear  viscosity and $n_0$ the density of the uniform fluid.
The data are overall
smooth up the highest packing fraction and we observe that the correction
to viscosity is relevant.
The presence of the crossover term determines a larger viscosity
over the entire range of packing fractions studied with respect to the
one obtained in the absence of it.

\begin{figure}[htb]
\begin{centering}
\includegraphics[clip,scale=0.6]{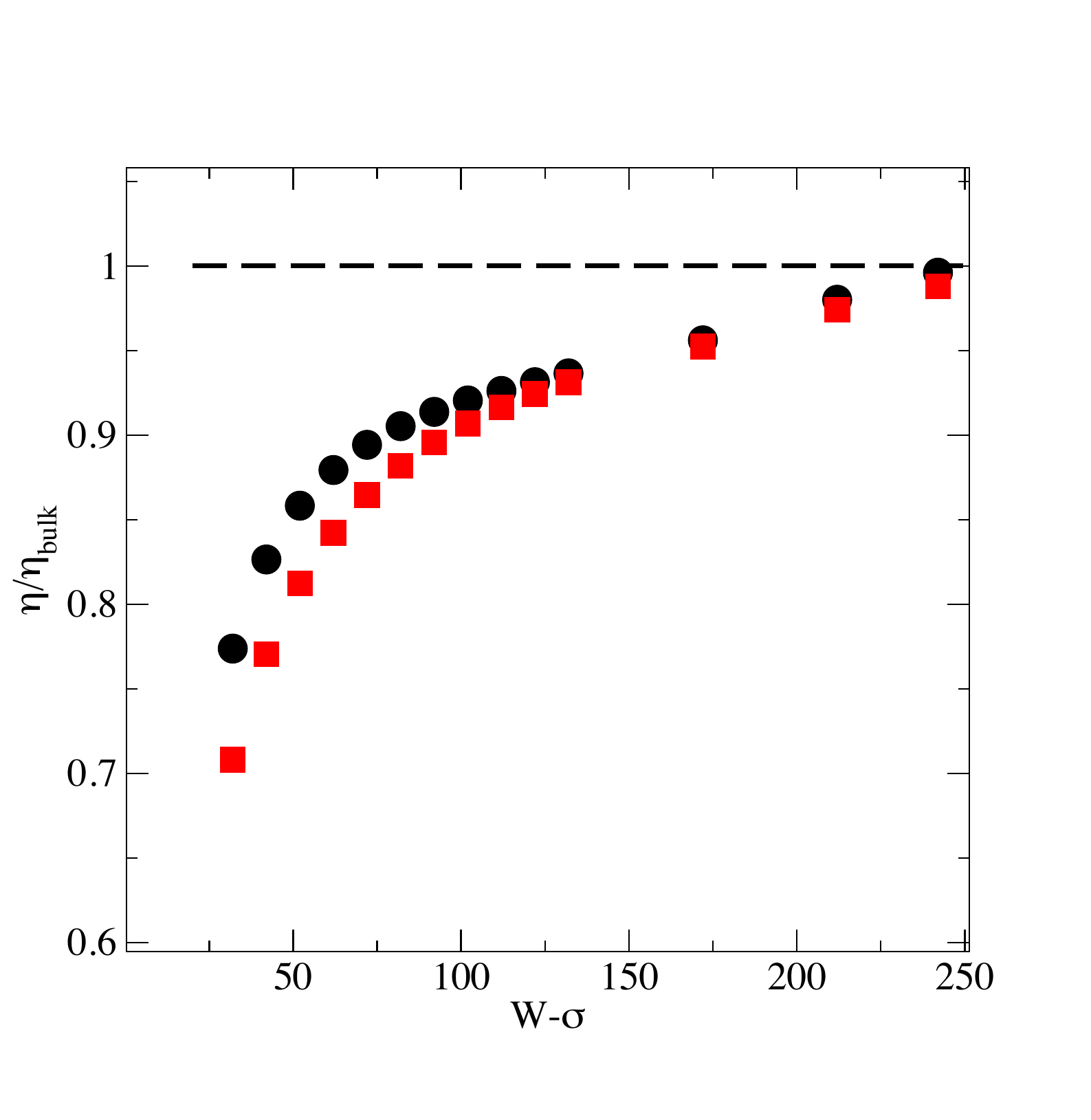} 
\caption{Normalized shear viscosity as a function of the effective slab width $(W-\sigma)$ expressed in lattice units.
Circles and squares correspond to packing fractions of $0.11$ and
$0.32$, respectively.}
\label{viscosityvsW} 
\end{centering}
\end{figure}

Next, we address the classical question regarding how the confinement
in a narrow space affects the viscosity of the fluid \cite{TravisGubbins,Todd1,Todd2}.
In our numerical tests, we consider the flow of a hard-sphere fluid
between two parallel plates and observe the density and velocity profile.
Since this a direct test of the viscosity of the fluid we expect a
change with respect to the previous version where the terms $A_{ij}$
and $B_{i}$ were not taken into account.
As shown in Fig.\ref{viscosityvsW} the measured viscosity converges 
for quite large channels of the order of $200$ lattice units towards
the predicted bulk value, whereas for small values of the width $W$ is consistently
lower, thus indicating that the walls strongly affect the motion of the adjacent layers.


\section{Conclusions}

\label{Conclusions}

In this study we have considered the corrections to the transport
coefficients which determine the cross terms between kinetic and collisional
contributions. At the price of including a couple of extra contributions in the term
describing the relaxation of fast modes, that is by lifting the BGK
approximation for these modes, we have obtained a new estimate for the
transport properties.
After deriving an explicit expression for these quantities,
we have evaluated the correction to viscosity via numerical simulations. It is
interesting to remark that the range of validity of the presented
LB algorithm does not extend to very low densities. The reason
being that, if one assumes a constant relaxation frequency $\omega_{0}$,
the value of the kinematic viscosity is too small 
and the small Knudsen number assumption breaks down.

The resulting form of the expression for the viscosity has similarities
with the one derived within the local average density model (LADM) by
Davis and coworkers \cite{Davis}. They assumed that the functional
dependence of transport coefficients on density can be computed via
the corresponding expressions in the Enskog theory of hard-sphere fluids,
by replacing the actual density $n(\rr,t)$ by the coarse
grained density, $\bar{n}(\rr,t)$. An approximation very similar
to the LADM can be derived within our method by using the explicit form
of $\Omega$. Such a recipe is in the same spirit as the early weighted
density density functional theories of equilibrium systems where the
local free energy density was assumed to be a function of $\bar{n}(\rr)$.
Moreover, as pointed out by Hoang and Galliero in a Molecular Dynamics study \cite{Galliero},
the LADM underestimates the importance of the local density as far
as the kinetic term is concerned. The present theory, does not suffer from
such a problem since it treats on a separate basis kinetic and potential
terms.

Finally, the presence of attractive interactions can be accommodated
in a mean field fashion by adding a Random Phase term in the evolution
equation. 
However, while representing a very important contribution to the equation of state, 
such a modification does not modify 
the transport coefficients (with the notable exception of the interspecies
diffusion coefficient in mixtures, see refs. \cite{KarkheckStell,JCP2011B}). 
The reason for this shortcoming is that in the Enskog-like equation
the details of the potential enter only through the cross-section.

\subsection*{Acknowledgments}

U.M.B.M. acknowledges the support of the Project COFIN-MIUR 2009.

\end{document}